\documentclass[preprint2]{aastex62}
\usepackage{epsfig}
\usepackage{natbib}
\usepackage{amsmath}
\usepackage{hyperref}

\newcommand{\kms}{km~s$^{-1}$}

\newcommand{\kmsMpc}{km~s$^{-1}$~Mpc$^{-1}$}

\begin{document}
\title{Cosmicflows-3: The South Pole Wall}

\author{Daniel Pomar\`ede}
\affil{Institut de Recherche sur les Lois Fondamentales de l'Univers, CEA Universit\'e Paris-Saclay, 91191 Gif-sur-Yvette, France}
\author{R. Brent Tully}
\affil{Institute for Astronomy, University of Hawaii, 2680 Woodlawn Drive, Honolulu, HI 96822, USA}
\author{Romain Graziani}
\affil{Laboratoire de Physique de 
Clermont, Université Clermont Auvergne, Aubière, France}
\author{H\'el\`ene M. Courtois}
\affil{University of Lyon, UCB Lyon 1, CNRS/IN2P3, IUF, IP2I Lyon, France}
\author{Yehuda Hoffman}
\affil{Racah Institute of Physics, Hebrew University, Jerusalem, 91904 Israel}
\author{J\'er\'emy Lezmy}
\affil{University of Lyon, UCB Lyon 1, CNRS/IN2P3, IUF, IP2I Lyon, France}

\begin{abstract}
Velocity and density field reconstructions of the volume of the universe within $0.05c$ derived from the {\it Cosmicflows-3} catalog of galaxy distances has revealed the presence of a filamentary structure extending across $\sim 0.11c$. The structure, at a characteristic redshift of 12,000~\kms, has a density peak coincident with the celestial South Pole.  This structure, the largest contiguous feature in the local volume and comparable to the Sloan Great Wall at half the distance, is given the name the South Pole Wall.
\end{abstract}

\bigskip

\smallskip
\section{Introduction}

The South Pole Wall rivals the Sloan Great Wall in extent, at a distance a factor two closer.  The iconic structures that have transformed our understanding of large scale structure have come from the observed distribution of galaxies assembled from redshift surveys: the Perseus$-$Pisces filament \citep{1982AJ.....87.1355G}, the CfA Great Wall \citep{1986ApJ...302L...1D}, the Sloan Great Wall \citep{2005ApJ...624..463G}.  The {\it Cosmicflows} program provides an alternative path for cosmographic studies.  The distribution of matter on large scales is inferred from the peculiar motions of galaxy test particles \citep{2012ApJ...744...43C, 2014Natur.513...71T, 2019ApJ...880...24T}.

The observed velocity of a galaxy $V_{obs}$ can be separated into the component due to the expansion of the universe $H_0 d$ and the residual line-of-sight peculiar velocity $V_{pec}$ with knowledge of the distance $d$ of the galaxy and the value of the Hubble constant $H_0$ compatible with the ensemble of measurements: to a first approximation $V_{pec} = V_{obs} - H_0 d$.  Although uncertainties with individual galaxies are large, the analysis benefits from the long range correlated nature of the cosmic flow, allowing the reconstruction of the 3D velocity field from noisy, finite and incomplete data \citep{1999ApJ...520..413Z}.  The current study draws on the {\it Cosmicflows-3} compilation of 17,669 distance measures \citep{2016AJ....152...50T} and the linear density reconstruction model of \citet{2019MNRAS.488.5438G}, supplemented by an alternative model by Hoffman et al. (in preparation).

\begin{figure*}[]
\plotone{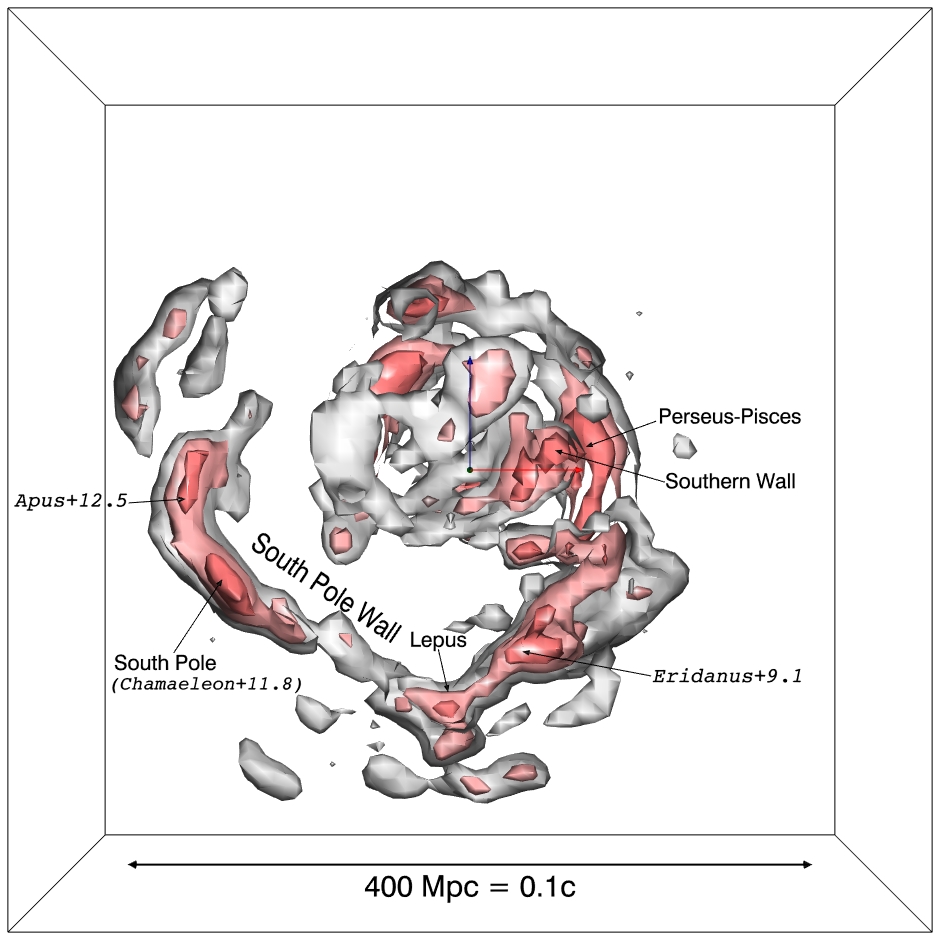}
\caption{A representation of overdensity structure in the Graziani et al. model of peculiar velocities inferred from {\it Cosmicflows-3} distances restricted to the supergalactic coordinate range $-13,000 < \rm{SGY} < 0$~\kms.  Iso-contours from grey to red are at density contrasts of $\delta=0.8$, 1.3, and 1.8. In this and ensuing figures, red, green, blue arrows emanating from the origin are 5,000~\kms\ in length along the positive SGX, SGY, SGZ axes respectively. The view is looking in from negative SGY.  Important features are labeled. }
\label{spw-y}
\end{figure*}

Our density reconstructions contain many fascinating structures.  Initial discussions highlighted voids \citep{2017ApJ...847L...6C,2019ApJ...880...24T}, chosen for study due to their proximity, relative simplicity, and because they have been given little attention.  This second look at the cosmographic features revealed in {\it Cosmicflows-3} reconstructions gives attention to a hitherto unknown dramatic filamentary structure.  The highest density part lies coincident with the celestial South Pole which justifies the name we give to the structure: the South Pole Wall.

\section{Density Reconstructions from {\it Cosmicflows-3} Distances}

The current discussion focuses on the model of peculiar velocities, ${\bf v}({\bf r})$, and densities $\delta({\bf r})$, described by \citet{2019MNRAS.488.5438G}.   In the assumed linear theory
\begin{equation}
\delta({\bf r}) = - \nabla \cdot {\bf v} / H_0 f
\label{eqdelta}
\end{equation}
where $f$ is the growth rate of structure given the $\Lambda$CDM parameters $\Omega_m=0.3$, $\Omega_{\Lambda}=0.7$, and $H_0=75$~\kmsMpc.  The Bayesian methodology seeks to derive the posterior probability of distances and the velocity field with nuisance parameters $h_{eff} = H_{optimal}/75$~\kmsMpc\ and a measure of nonlinear effects, $\sigma_{NL}$, given as data the observed distance moduli, $\mu_i$, redshifts, $z_i$, and estimated errors.  Sampling from the model and data constraints gives the mean and standard deviation of the linear velocity field at each location within $z \sim 0.05$.   The Gibbs sampling method utilizes a Markov Chain Monte Carlo procedure \citep{2016MNRAS.457..172L}. Details are described by \citet{2019MNRAS.488.5438G}.  The analysis results in the specification $h_{eff} = 1.02 \pm0.01$.  The model is saved on a three-dimensional grid with 6.25~Mpc spacings assuming the fiducial value of $H_0=75$~\kmsMpc.

As a check, an independent analysis based on the same {\it Cosmicflows-3} distance information has been carried out with the Wiener Filter with Constrained Realizations methodology employed with earlier {\it Cosmicflows-2} studies \citep{1991ApJ...380L...5H, 1999ApJ...520..413Z, 2012ApJ...744...43C, 2014Natur.513...71T}.  The results of the independent study will be summarily presented as providing confirmation of the essential claims made in this paper.  A full description of the Wiener Filter/Constrained Realizations model with a new treatment of biases is in preparation (Hoffman et al. 2020).

\begin{figure*}[]
\includegraphics[scale=0.27]{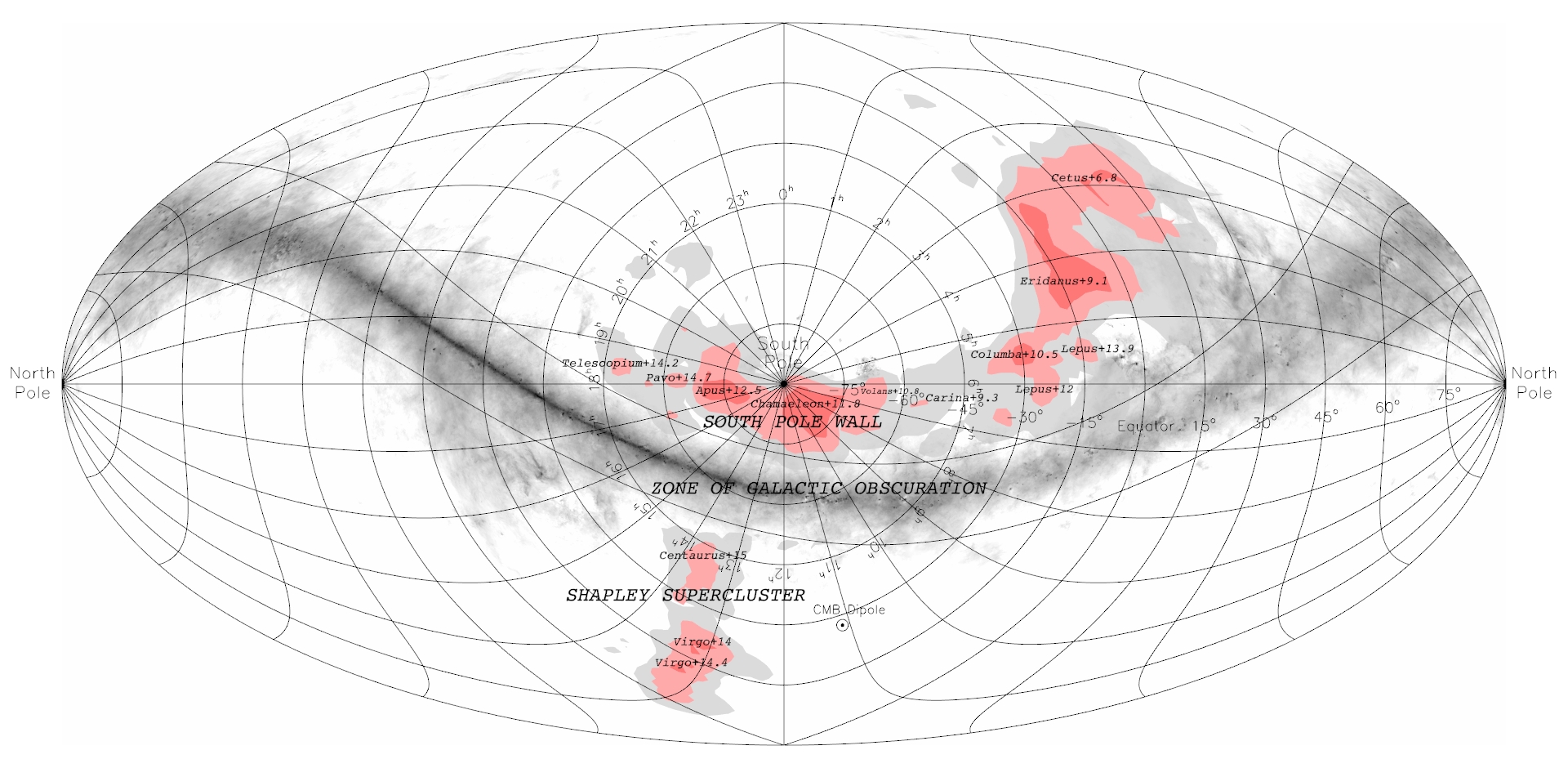}
\caption{A projection of the South Pole Wall in celestial coordinates.  A major feature of the wall lies in the direction of the terrestrial South Pole.   The plane of the Milky Way is shown by a dust map in shades of grey. The curves that parallel the plane are at $b = \pm 10^{\circ}$. Major features are labeled, including the proximity of the Shapley Supercluster on the opposite side of the zone of obscuration and the apex of the cosmic microwave background temperature dipole.
}
\label{cartesian}
\end{figure*}

\section{The South Pole Wall}

The structure to be described lies south of the plane of the Milky Way in galactic coordinates and predominantly in the celestial south.  In supergalactic coordinates, the galactic south lies at negative SGY.  In order to eliminate confusion from the galactic north, the overdensities in the \citet{2019MNRAS.488.5438G} model are shown in Figure~\ref{spw-y} restricted to SGY$<0$, with the viewer looking in from $-$SGY.  There is a secondary cut requiring SGY$>-13,000$~\kms\ to eliminate projection confusion. An obvious continuous structure runs from the feature labeled Apus+12.5\footnote{We follow a naming convention \citep{2019ApJ...880...24T} based on the projected constellation of the feature and redshift in units of 1000~\kms, positive if an overdensity and negative if an underdensity.} to Lepus and from there through the Funnel (Eridanus+9.1) to Perseus$-$Pisces and the Southern Wall.  

The more celestially northerly and nearby of these features are relatively well known.  The Perseus$-$Pisces filament has long been established as a prominent structure \citep{1982AJ.....87.1355G, 1988lsmu.book...31H}.  Likewise, the Southern Wall was identified early in studies of large scale structure \citep{1990AJ.....99..751P}.  The Funnel is a name given to describe the gathering of flow patterns emanating from Perseus$-$Pisces and the Southern Wall that proceed toward Lepus \citep{2017ApJ...845...55P}. This latter publication was based on a study using {\it Cosmicflows-2} distances.  In that earlier study the flow patterns upon reaching the already distant Lepus region passed through {\it astrus incognito} to reach the Shapley concentration of galaxies \citep{1989Natur.338..562S, 1989Natur.342..251R}.  It has taken the more plentiful {\it Cosmicflows-3} compilation to edify the nature of structure beyond Lepus.

It is seen in Figure~\ref{spw-y} that there is a pronounced continuous feature running from Lepus to a high density feature labeled Apus+12.5. Chamaeleon+11.8 is the highest density peak along this extended filament, coincident with the celestial South Pole.  This coincidence is shown in Figure~\ref{cartesian}, a plot in celestial coordinates centered on the South Pole.  

The regions of obscuring dust of the Milky Way are represented in this second figure.  The South Pole Wall as we identify it lies entirely to the galactic south of the Milky Way, but hugs close to the galactic plane.  The proximity of the populous Shapley overdensity across the zone of obscuration from the South Pole Wall is to be noted.  The apex of the cosmic microwave background dipole temperature fluctuation is nearby.  Our models based on peculiar velocity flows recover the density peak at the South Pole even though the direction is obscured by the Chamaeleon molecular cloud complex \citep{1962ZA.....55..290H}.

\begin{figure*}[]
\includegraphics[scale=0.365]{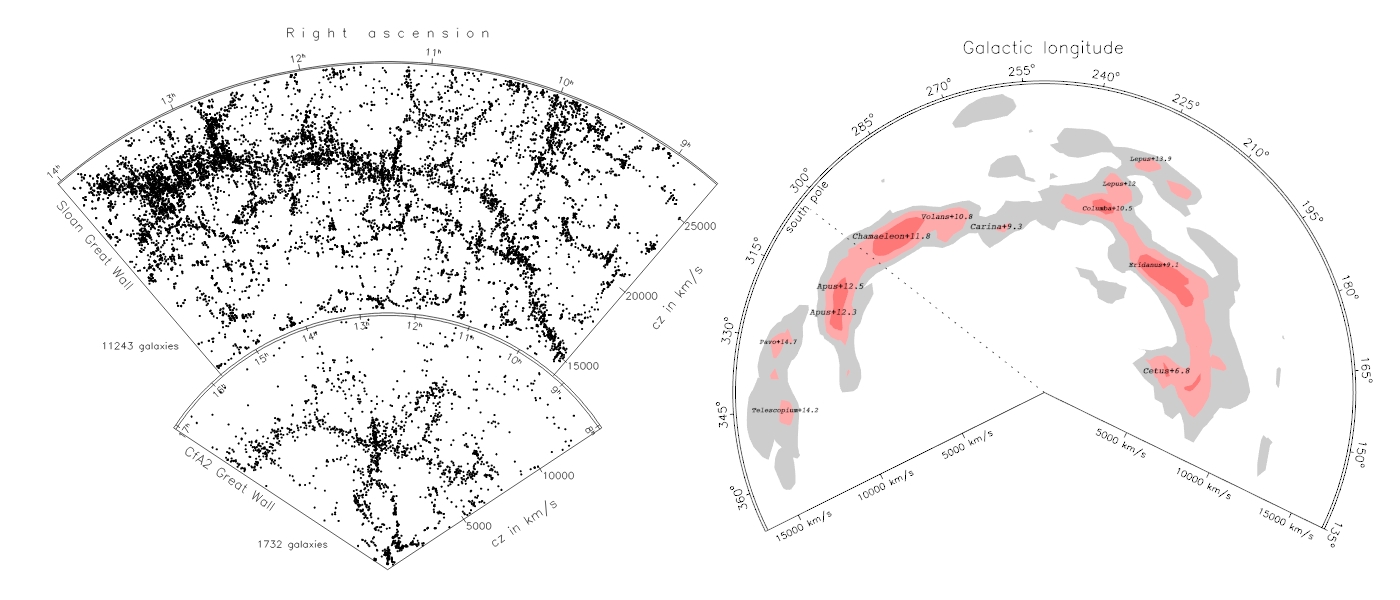}
\caption{A comparison of the extent of the South Pole Wall with two other nearby structures: the CFA Great Wall (bottom left) and the Sloan Great Wall (top left).  The South Pole Wall at the right is represented by isodensity contour levels derived from the peculiar velocity field.  The linear scales are the same.  The South Pole Wall is roughly intermediate in distance from us compared with the other two.
}
\label{3walls}
\end{figure*}

The South Pole Wall extends across a wide swath of the sky, especially if taken to continue to the region of the Funnel.  Since it adheres reasonably closely to a constant latitude with respect to the galactic plane, it is conveniently mapped in galactic longitude against redshift as seen in the right panel of Figure~\ref{3walls}.  The left panel is a reproduction of a rather famous illustration of the Sloan Great Wall and CFA Great Wall \citep{2005ApJ...624..463G} for comparison.  The extent of the South Pole and Sloan walls are comparable.  The run from Apus to Lepus over $\sim 98^{\circ}$ at latitude $-20^{\circ}$ and $\sim 12,000$~\kms\ covers $\sim 19,000$~\kms $\sim 250$~Mpc. The bend toward us inward to $\sim 7000$~\kms\ over $\sim 85^{\circ}$ from Lepus to the Funnel (Cetus+6.8) is of length $\sim 13,000$~\kms $\sim 170$~Mpc.  End-to-end, from Apus to the Funnel is $\sim 0.11c \sim 420$~Mpc.  This structure is a factor of two {\it closer} that the Sloan Great Wall, evidence that such large contiguous entities are not unusual.

\begin{figure*}[]
\includegraphics[scale=0.408]{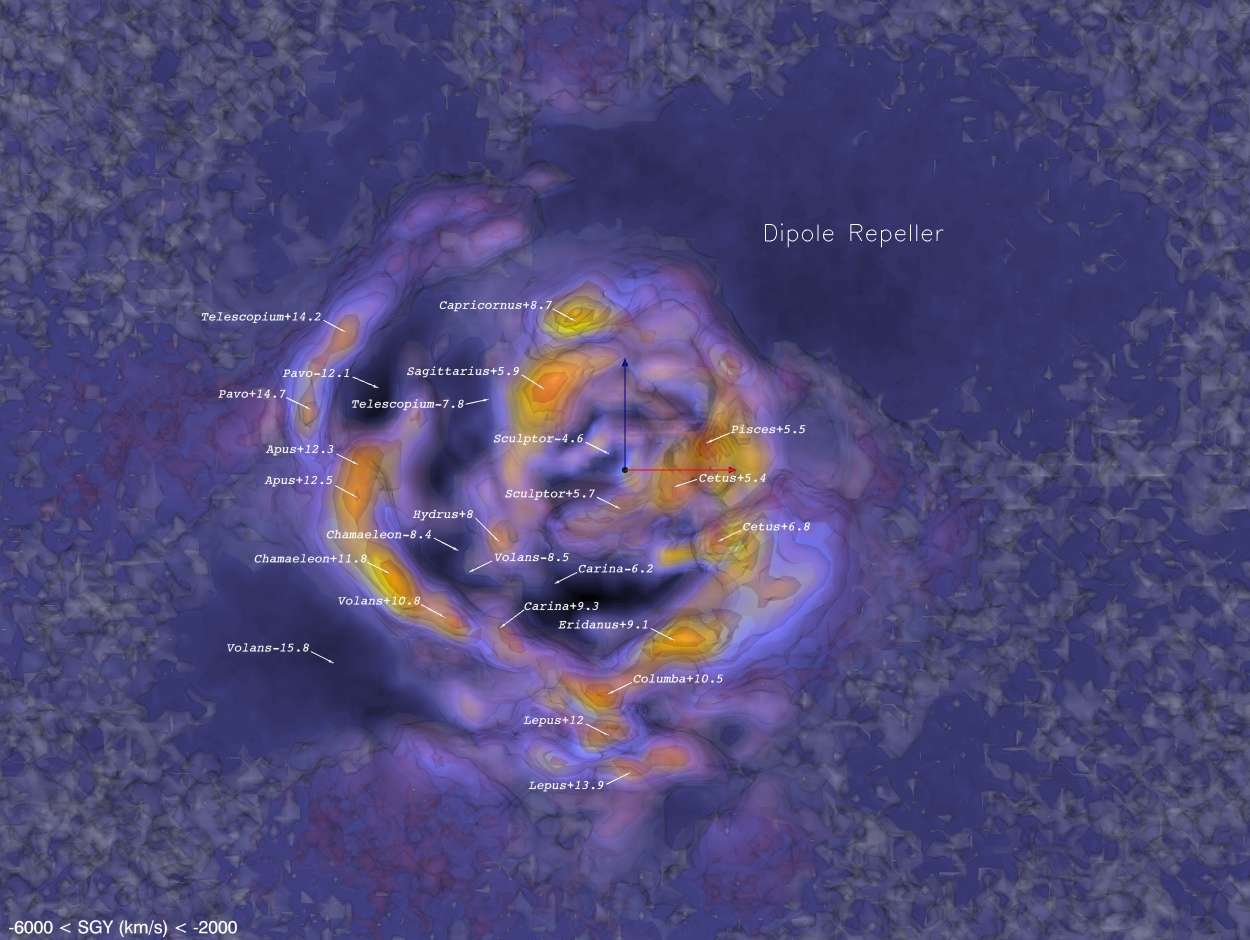}
\caption{An alternate representation of overdensity structure in the Graziani et al. model restricted to the supergalactic coordinate range $-6,000 < \rm{SGY} < -2,000$~\kms.  The view is looking in from negative SGY.  Important features are labeled using as the naming convention the constellation in the line of sight and the redshift in units of 1000~\kms.  Positive redshift values denote overdensities and negative redshift values denote underdensities.}
\label{spw-y_names}
\end{figure*}

\begin{figure*}[]
\plotone{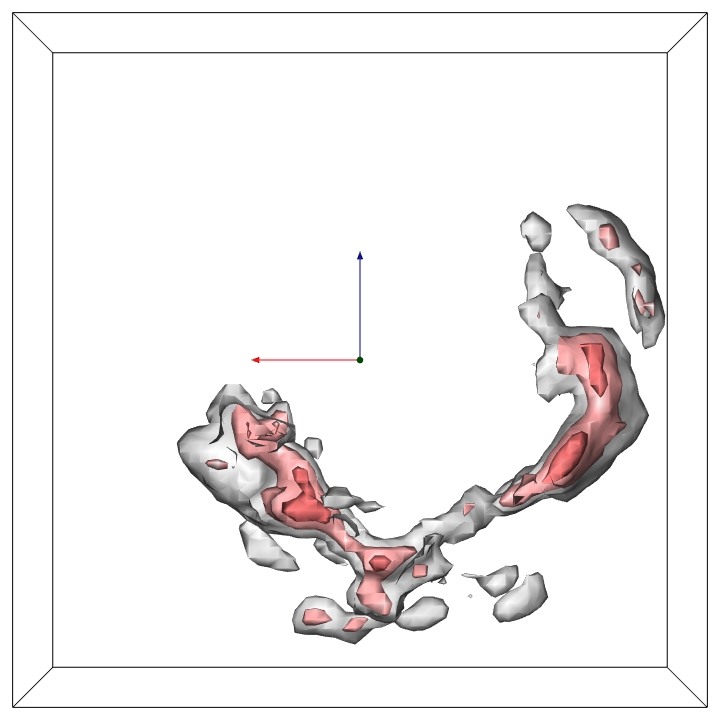}
\caption{Isodensity contours derived from the observed peculiar velocities of galaxies in the {\it Cosmicflows-3} sample of distances.  The South Pole Wall is seen in isolation in a view looking in along the positive SGY axis in supergalactic coordinates.  As in other figures, the red and blue arrows are each 5,000~\kms\ in length, directed toward +SGX and +SGZ respectively. An  \href{https://sketchfab.com/3d-models/656f5663266c4d2d914e0fdbd1237022}{interactive model} is available online.
}
\label{spw+y}
\end{figure*}

\begin{figure*}[]
\plotone{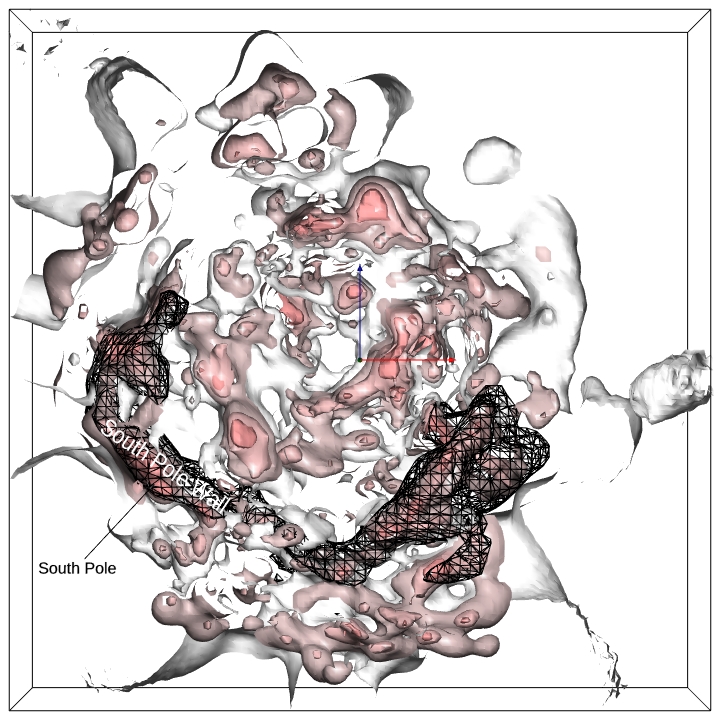}
\caption{A representation of overdensity structure in the Hoffman et al. model of peculiar velocities inferred from {\it Cosmicflows-3} distances restricted to the supergalactic coordinate range $-6,000 < \rm{SGY} < -3,000$~\kms.  The view is looking in from negative SGY, using four solid isosurfaces: delta=0 in grey, delta=0.5, 1.0, and 1.5 in nuances of red. For comparison, the South Pole Wall as reconstructed using the Graziani et al (2019) model is superimposed as a black wireframe isosurface. An interactive visualization of the same objects can be \href{https://sketchfab.com/3d-models/cf3bgc600-256-4isos-sgy-6000a-3000-vs-spwrg-v002-0d5c7fecea4f49cf94e1552dc29c5c3c}{viewed online}.  
}
\label{spw-y_YH}
\end{figure*}

Figure~\ref{spw-y_names} is another view of the \citet{2019MNRAS.488.5438G} model with a tighter slice, only 4,000~\kms\ thick, and including negative contours, again looking in from along the negative SGY axis.
This model results from a Gibbs sampling MCMC chain of 2633 constrained realizations. Chain variances in the vicinity of the South Pole Wall are $\sigma_\delta \simeq 1.65$.    The most tenuous part of the South Pole Wall adjacent Carina+9.3 attains $\delta \sim 1$, a robust value. Local uncertainties in the composite $\delta$ are estimated to be $\sim 0.1$.

Figure~\ref{spw+y} shows the full extent of the posited South Pole Wall stripped of features that we consider unrelated.  Now the view is from along the positive SGY axis, with arrows 5,000~\kms\ in length directed from the supergalactic origin toward +SGX (red) and +SGZ (blue).  This figure provides an entry point into a Sketchfab viewer allowing an interactive visualization of the model.\footnote{
Sketchfab viewers allow immersive, interactive navigation within 3D models. Use mouse action for rotation, zoom, translation of the view point.  Selected features are indicated by the means of numbered annotations. An Autopilot mode guides you through the corresponding predefined stations.}  SGY boundaries are set at zero and $-13,000$~\kms. The overdensity contours are again at $\delta = 0.8$, 1.3, and 1.8.

It has become evident that, wherever there are substantial overdensities, adjacent there are voids.  The South Pole Wall conforms to this expectation.  To the near side lies the Sculptor Void, appropriately the most prominent void within the {\it Cosmicflows-3} study volume \citep{2019ApJ...880...24T}.  At least to a partial degree on the far side lies the poorly documented Eridanus Void.  In Fig.~\ref{spw-y_names} the abysses Carina-6.2, Volans-8.5 and Chamaeleon-8.4 are density minima within the Sculptor Void.  \citet{2019ApJ...880...24T} provide a video that illustrates the domain of the Sculptor Void and the proximity of the South Pole Wall as a bounding feature.

The Wiener Filter/Constrained Realization model of the {\it Cosmicflows-3} peculiar velocity observations of Hoffman et al. (2020, in preparation) provides the map of overdensities shown in Figure~\ref{spw-y_YH}. The slice is 3,000~\kms\ thick, looking in from $-$SGY.
The resolution is twice as high as that of \citet{2019MNRAS.488.5438G} with grid spacings of 3.1~Mpc with fiducial $H_0 = 75$~\kmsMpc.  The structure tends to break up into a lumpier distribution as a consequence of the higher resolution.  The continuous structure from Apus to the Funnel is seen, albeit with greater confusion.  A wireframe representation of the structure shown in Fig.~\ref{spw+y} from the Graziani et al. model is superimposed.
An interactive Sketchfab model can be accessed from the caption for Fig.~\ref{spw-y_YH} that provides better clarity than afforded by the static view.
Full discussion of the complex structure found from the Hoffman et al. analysis awaits publication of this alternative model of {\it Cosmicflows-3} peculiar velocities.

\begin{figure*}[]
\plotone{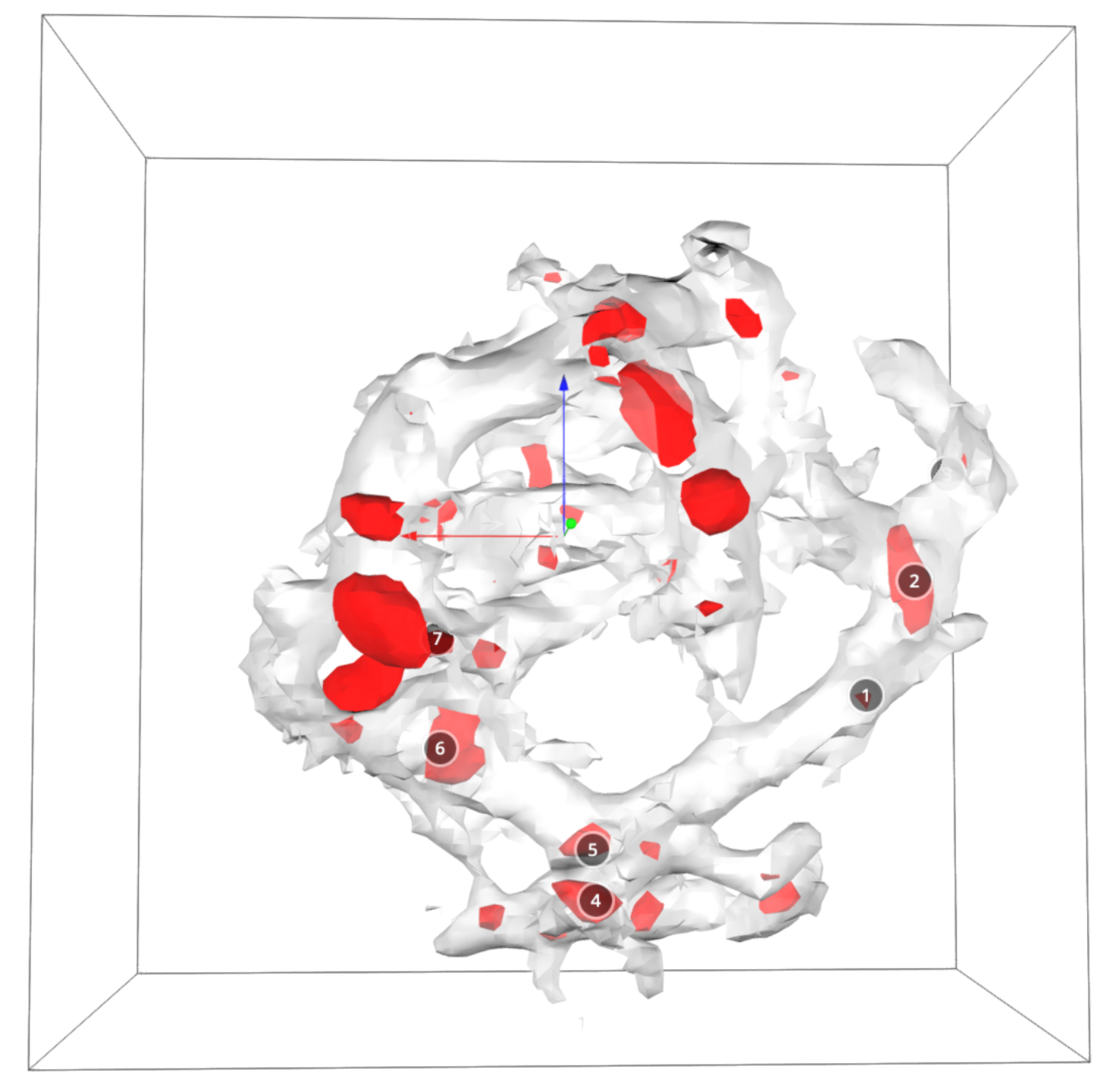}
\caption{A V-web representation of structure based on the Graziani et al. model of the {\it Cosmicflows-3} peculiar velocity field.  Knots are represented in red and filaments in off-white.  The display is restricted to $-13,000 < {\rm SGY} < 0$~\kms.   The three-dimensional structure can be explored with the \underline{\href{https://sketchfab.com/3d-models/d3ee5ee15ed34f0aacecc1f6a8a8124c}{online interactive model}}.
}
\label{V-web}
\end{figure*}

As an alternative to the relative density definition of structure following from Eq.~\ref{eqdelta}, knots, filaments, sheets, and voids can be defined through the determination of velocity shear at each voxel \citep{2012MNRAS.425.2049H, 2017ApJ...845...55P}: 
\begin{equation}
\Sigma_{\alpha\beta} = -  ( \partial_\alpha v_\beta  +  \partial_\beta v_\alpha)/2H_0
\label{eq;vweb}
\end{equation}
where partial derivatives of the velocity ${\bf v}$ are determined along directions $\alpha$ and $\beta$ of orthogonal Cartesian axes, normalized by the average expansion given by the Hubble Constant, $H_0$.  Eigenvalues indicating collapse have negative values. Knots are characterized by three eigenvalues of collapse, voids by three eigenvalues of expansion, and filaments and sheets are the intermediate cases.  Although not easy to follow in a static view such as the representation on a page, the South Pole Wall is easy to discern 
in the interactive Figure~\ref{V-web},
which gives access to the capabilities of rotation and zoom.
There is impressive continuity of the South Pole Wall as a filament as defined by the quantitative description provided by the eigenvalues of local velocity shear.  The eigenvalues specify that there are knots embedded within the filament at Apus+12.9, Chamaeleon+11.8, Lepus+12, Columba+11.1, and Eridanus+10.2.  The alternative Wiener filter model to be reported by Hoffman et al., with lower signal to noise at the redshift of the South Pole Wall, identifies the run of the structure as a wall or sheet from an eigenfunction analysis.

The video map introduced  in the summary section, shows the V-web representation at t=4min05s. The video provides a visual comparison with
individual galaxies at their redshift positions drawn from the 2MASS Extended Source Catalog \citep{2000AJ....119.2498J}.  In spite of the substantial redshift of the South Pole Wall and the flirtation with the zone of obscuration, there is a substantial positive association of galaxies from the redshift catalog and the boundaries of the South Pole Wall. 

In hindsight, the South Pole Wall can be recovered from the projection maps given with the final release of the 6dF Galaxy Survey \citep{2009MNRAS.399..683J}.  In the declination slices of Fig.~8 in that paper, the structure is represented in all 6 declination intervals at $\sim 10-12,000$~\kms, rotating counter-clockwise from about 1 o'clock in the most equatorial slice, through the features labeled Columba-Lepus and Puppis-Carina at 12 0'clock, then seen in rotation all the way to 6 o'clock in the slice south of declination $-60^{\circ}$.  Regrettably, this southernmost slice is poorly covered because of missing tiles in the 6dF survey.  Nonetheless, much of the run of the wall can be seen in the top panel of Fig.~1 of \citet{2009MNRAS.399..683J} running down from the concentration labelled Eridanus-Lepus and, better, in the lower panel of their Fig.~6, the feature in green and yellow running along latitude $b=-30^{\circ}$ from $l\sim210^{\circ}$ to where data thins out near the south celestial pole at $l\sim 280^{\circ}$.

\section{Flow Streamlines}
\label{sec:flows}

\begin{figure*}[]
\includegraphics[scale=0.28]{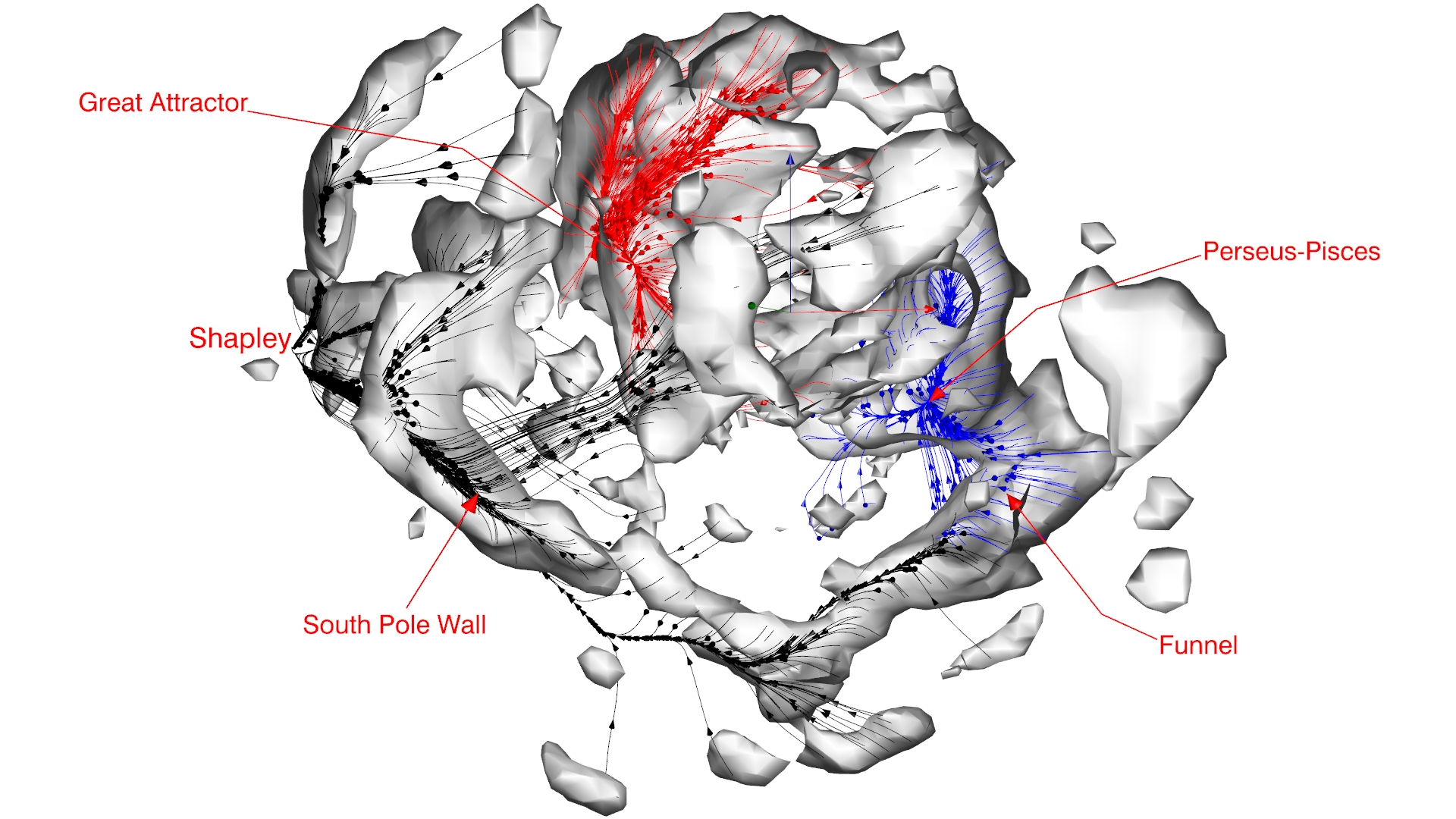}
\caption{Flow streamlines seeded within the density unity contour of the Graziani et al. model.  Flowlines proceed from seed position to one of three accumulation points associated respectively with the Shapley concentration, the Perseus$-$Pisces filament, and the Great Attractor.  Flow lines associated with seeds along most of the South Pole Wall proceed to the Shapley concentration. There is a divergence between flows toward Shapley and flows toward Perseus$-$Pisces in the region of the Funnel indicated by an arrow. Details of the flow lines are best appreciated by opening the \href{https://sketchfab.com/3d-models/0754e27c1075455cae090f1038d3cfbc}{interactive model}.
}
\label{flows}
\end{figure*}

Yet another way of unveiling the large scale structures is through the continuity of flow streamline patterns \citep{2017NatAs...1E..36H}. The displacements of individual galaxies is tiny on the scale of large scale structures, only of order a few Mpc over the age of the universe.  The expansion of the universe completely dominates over peculiar velocities on scales larger than several Mpc.  Nevertheless, streamlines are useful constructs for the delineation of separate gravitational basins \citep{2019MNRAS.489L...1D}.  

Streamlines are trajectories in space whose tangents coincide with the direction of the velocity field at any point.  Starting from a seed point $\vec{r}$ a streamline is defined by $\vec{r}=\int_{\vec{r}_0}^{\vec{r}} \vec{v}\cdot d\vec{s}$ where $d\vec{s}$ is the line element of the trajectory.  Streamlines can be seeded in an ordered manner or randomly.  If integrated long enough they either converge at the attractors of the flow field or escape out of the computational box.
 \citep{2001astro.ph..2190H}.  A multitude of seeds will generate a multitude of streamlines that converge on the local extrema of the velocity potential.  These are the stationary knots of the V-web.

In the current application, seeds are located within regions of overdensity $\delta>0.9$.  The consequent flow lines are seen in Figure~\ref{flows} and, even better, in the associated interactive model. 
It will be seen that there are three cumulation points for streamlines, identified in turn with the general region of the Great Attractor \citep{1987ApJ...313L..37D} at the heart of Laniakea Supercluster \citep{2014Natur.513...71T}, the Perseus$-$Pisces filament \citep{1988lsmu.book...31H}, and the Shapley concentration {\citep{1989Natur.338..562S, 1989Natur.342..251R}.  Essentially all the seeds placed along the South Pole Wall terminate in the Shapley concentration.  The only exceptions are those at the extreme northern end in the region being referred to as the Funnel. Here there is a discontinuity, with streamlines flowing instead to the Perseus$-$Pisces basin.

This pattern of flows along filaments with divergent points between adjacent gravitational basins is becoming familiar as mapping of velocity fields become more robust.  \citet{2017ApJ...845...55P} illustrate several such cases along filaments connecting Laniakea and Perseus$-$Pisces; specifically the structures called the Centaurus$-$Puppis$-$PP filament and the Centaurus$-$Virgo$-$PP filament, so-named because of the filament end points and the routes that they take.

In the case of the South Pole Wall the structure is immensely larger but the characteristic is similar.  Just as seen in simulations, filaments interconnect across the cosmic web. Velocity field information reveals how these structure shear between adjacent knots of the V-web.

\section{Dynamical Influence}

The location of the apex of the cosmic microwave background dipole is indicated in Fig.~\ref{cartesian}.  The Shapley Supercluster lies nearby, long a suspected major player in the generation of the 630~\kms\ motion of the Local Group with respect to the cosmic rest frame \citep{1989Natur.338..562S, 1989Natur.342..251R}.  In detail, this motion has arisen out of innumerable peaks and valleys in the distribution of matter on a wide range of scales.  The mass concentration from Centaurus to Norma clusters at the heart of Laniakea Supercluster must be important \citep{1988ApJ...326...19L}.  The significant pull and push contributions of the nearby Virgo Cluster and Local Void have received recent emphasis \citep{2019ApJ...880...24T}.  On a very large scale, there is the push provided by the Dipole Repeller \citep{2017NatAs...1E..36H}.

Given the proximity in direction of the densest part of the South Pole Wall, it is to be asked if this feature also plays a significant attractive roll in the motion reflected in the CMB dipole.  The attraction at our position of the overdensity within a volume running from the peaks labeled Volans+10.8 to Telescopium+14.2 in Fig.~\ref{3walls} gives rise to a velocity of $42\pm11$~\kms.  By comparison, a volume enclosing the Shapley concentration is pulling at $51\pm12$~\kms.\footnote{The alternative Wiener Filter model gives an attractive pull by Shapley of $67\pm27$~\kms \citep{2017NatAs...1E..36H}.}  The influences are remarkably comparable.  The densities of galaxies and modelled mass are much higher in Shapley and, indeed, the pattern of galaxy motions, discussed in \S \ref{sec:flows}, flow through the South Pole Wall toward a terminus in Shapley, substantiating the primacy of Shapley.

\section{Summary}

\begin{figure*}[]
\plotone{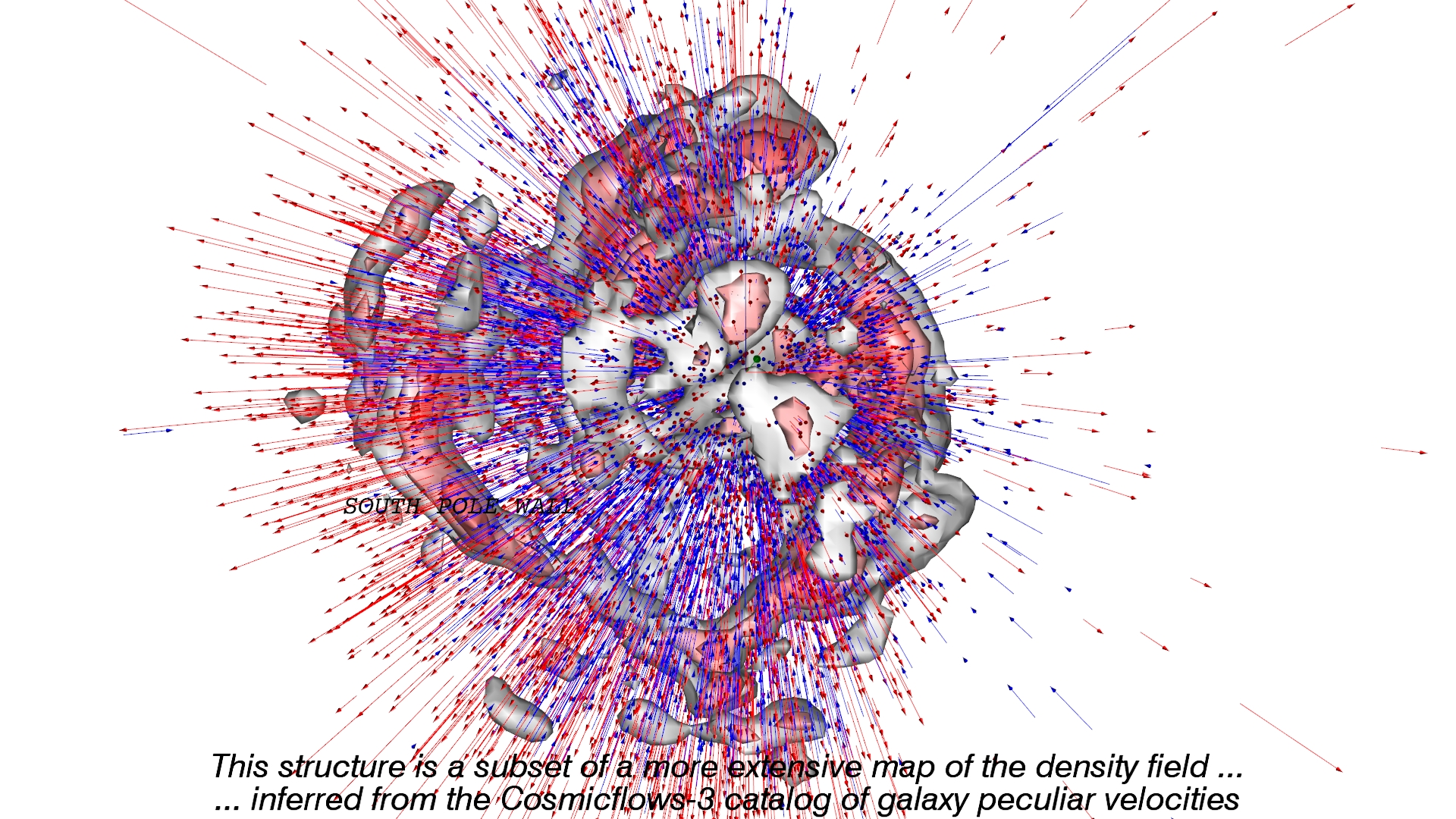}
\caption{Entry to video description of the South Pole Wall. The video in high-resolution is available online at \href{https://vimeo.com/389251832/6feeec421f}{https://vimeo.com/389251832/6feeec421f}. In this image, the radially directed arrows give representation to {\it Cosmicflows-3} peculiar velocities by their length and color (red outward and blue inward).  The modeled density field is given representation by the iso-density surfaces.} 
\label{video}
\end{figure*}

The accompanying 5 minute video (see interactive Figure~\ref{video}) 
encapsulates the salient points of this discussion.  Observational limitations must be acknowledged.  Recognition of the South Pole Wall feature has only been possible because of the contribution of the Six Degree Field Galaxy Survey component of {\it Cosmicflows-3} \citep{2014MNRAS.445.2677S}. The redshift limit of this contribution is 16,000~\kms.  The South Pole Wall as we constitute it walks a constrained line inside this limit at $\sim 13,000$~\kms\ and, at $b \sim -20^{\circ}$, the other observational impediment of the zone of obscuration of the Milky Way.  

The proximity of the Shapley concentration of rich clusters with similar velocities \citep{1989Natur.338..562S, 1989Natur.342..251R} and the direction of our motion inferred from the cosmic microwave background dipole \citep{1996ApJ...473..576F} both immediately to the north of the galactic plane begs the question of what we might be missing.  Resolution will require numerous and accurate distance measures to significantly larger redshifts.

In addition to these redshift and obscuration edge effects, there are other ambiguities regarding the full extent of the South Pole Wall.  The $\sim 19,000$~\kms\ run of a rather straight filament from Apus through the celestial South Pole to Lepus is most striking. Then there is the $\sim 13,000$~\kms\ long complex between Lepus and the Funnel after a bend in direction at Lepus.  Should these structures be considered conjoint?  We suggest yes.  It is in the Funnel that our seeded streamlines shear between flows toward Perseus$-$Pisces and Shapley.  From the Funnel, all along the posited South Pole Wall, streamlines flow along this structure before jumping across the zone of obscuration to reach the Shapley concentration. 
The distance between Lepus and Shapley is 19,000~\kms$\sim 250$~Mpc.

Given the limitations imposed by the boundaries of our study, where does the connectivity end?  There is a hint of a related feature at 18,000~\kms\ in the constellation Vela in the zone of obscuration \citep{2019MNRAS.490L..57C}. Indeed, perhaps we should be informed by simulations \citep{2005Natur.435..629S} that the cosmic network of overdensities is quasi-connected over indefinite lengths. 

In any event, the reasonable case can be made that the South Pole Wall has an extent of at least $0.11c \sim 420$~Mpc.  This extent is impressive given that the effective radius of the {\it Cosmicflows-3} compilation of distances is $0.05c$.  The South Pole Wall is the largest contiguous structure within this volume.  We will not be certain of its full extent, nor whether it is unusual, until we map the universe on a significantly grander scale.

\bigskip

\noindent
{\bf Acknowledgements}

Financial support for the {\it Cosmicflows} program has been provided by the US National Science Foundation award AST09-08846, an award from the Jet Propulsion Lab for observations with {\it Spitzer Space Telescope}, and NASA award NNX12AE70G for analysis of data from the {\it Wide-field Infrared Survey Explorer}.  HC acknowledges support from the Institut Universitaire de France, the CNES, and
the Project IDEXLYON at the University of Lyon under the Investments for the Future Program (ANR-16-IDEX-0005). YH is supported by the Israel Science Foundation grant ISF 1358/18.
  
\bibliography{spw}
\bibliographystyle{aasjournal}

\end{document}